\documentclass[ iop]{emulateapj}
\usepackage{amsmath}
\usepackage{amsfonts}
\usepackage{graphicx}
\usepackage{xcolor}

\usepackage{mathtools}

\makeatletter
\newcommand\Label[1]{&\refstepcounter{equation}(\theequation)\ltx@label{#1}&}
\makeatother

\usepackage[caption=false]{subfig}

\newcommand{\beq}{\begin{equation}}
\newcommand{\eeq}{\end{equation}}
\newcommand{\bea}{\begin{eqnarray}}
\newcommand{\eea}{\end{eqnarray}}
\newcommand{\mrm}{\mathrm}

\newcommand{\pt}{\partial}

\newcommand{\Lwmax}{L_{{\rm max}}}
\newcommand{\Rsf}{R_{\rm sf}} 
\newcommand{\Hprad}{H_{p,{\rm rad}}}
\newcommand{\omegarad}{\omega_{\rm rad}}

\shortauthors{Ro \& Matzner}

\begin{document}

\title{Shock Dynamics in Stellar Outbursts: I. Shock formation} 

\author{Stephen Ro\email{ro@astro.utoronto.ca} \& Christopher D. Matzner}
\affil{Department of Astronomy \& Astrophysics, University of Toronto, 50 St. George St., Toronto, ON M5S 3H4, Canada}

\begin{abstract}
Wave-driven outflows and non-disruptive explosions have been implicated in pre-supernova outbursts, supernova impostors, LBV eruptions, and some narrow-line and superluminous supernovae. To model these events, we investigate the dynamics of 
stars set in motion by strong acoustic pulses and wave trains, focusing here on nonlinear wave propagation, shock formation, and an early phase of the development of a weak shock.   We identify the shock formation radius, showing that a heuristic estimate based on crossing characteristics matches an exact expansion around the wave front and verifying both with numerical experiments.  Our general analytical condition for shock formation applies to one-dimensional motions within any static environment, including both eruptions and implosions, and can easily be extended to non-stationary flows.  We also consider the early phase of shock energy dissipation.  We find that waves of super-Eddington acoustic luminosity always create shocks, rather than damping by radiative diffusion.  Therefore, shock formation is integral to super-Eddington outbursts.
\end{abstract}

\section{Introduction}

Observational examples of progidious pre-supernova (pre-SN) mass loss now abound. Supernova impostors  \citep{2012ASSL..384..249V} are re-classified as intense luminous blue variable (LBV) outbursts once their progenitors are seen to survive.  Some, like SN 2006jc, SN 2009ip, SN 2015bh, and LSQ13zm, undergo one or more eruptions before terminal explosion 
\citep{2007Natur.447..829P, 2007ApJ...657L.105F, Margutti14_Panchromatic09ip, 2014MNRAS.438.1191S,2016arXiv160609025T,2016MNRAS.459.1039T}.   A few percent of core-collapse SNe exhibit narrow lines from dense circumstellar interaction, indicating intense phases of pre-SN mass loss \citep{2011MNRAS.412.1522S,2012ApJ...744...10K,2014ARA&A..52..487S,2014MNRAS.439.2917M}. \cite{Ofek14_PrecursorsToIInAreCommon} find that precursors are common in hydrogen-rich, narrow-line (type IIn) SNe, and \cite{2016arXiv160106806M} find evidence for outbursts ejecting $\sim 1\,M_\odot$  in SN 2014C and similar events prior to $\sim$10\% of type Ibc SNe. 

It is not always clear whether each mass-loss episode results from a single shock-driven outburst, an extended wind, or both. However, intense mass-loss is expected in a pre-SN stellar evolution for both low and high progenitor masses. In the low-mass ($\sim 9-11 M_\odot$) progenitors of electron-capture SNe, oxygen and silicon shell burning releases a sequence of pulses, building from $10^{49}$ to $10^{50}$ ergs over the last year of the star's life, allowing for the possible ejection of the stellar envelope \citep{2015ApJ...810...34W}.   At an order of magnitude higher initial mass ($\sim 95-130 M_\odot$), pulsational pair  instability is expected to eject a series of massive shells \citep{2007Natur.450..390W}. Outside these mass windows, \citealt{2012MNRAS.423L..92Q}, \citealt{2014ApJ...780...96S}, and \cite{2014ApJ...785...82S} argue that the enhanced pre-SN mass loss is driven by waves excited in zones of vigorous convection.

Waves deposit their energy by either radiative damping or shock dissipation. Super-Eddington rates of acoustic dissipation can stimulate intense winds like those seen from LBVs and Type IIn SN progenitors. \citet{2016MNRAS.458.1214Q} envision radiative diffusion as the dominant form of wave dissipation in such events, but as we shall see, shock dissipation is more relevant. Moreover, the existence of $\sim$5000\,km/s motions around $\eta$\,Carinae \citep{2008Natur.455..201S} implies shock driving, as do 2000-7000\,km/s speeds in the 2009ip precursor \citep{2011ApJ...732...32F}. 

These considerations motivate a detailed investigation of shocks within stars, which we begin here by analyzing the birth and early phase of a radially-propagating shock front. \cite{2010MNRAS.405.2113D} has noted that shocks may be responsible for many types of outbursts, and that shocks occur naturally when energy is released over a period shorter than the dynamical time.  However, energy is usually released deep within a star where sound speeds are relatively large, so part of the deposited energy must first travel outward as a sound pulse or continuous wave.  If the sound is sufficiently intense, it will convert into a shock at some point within the star.  
Indeed, shocks are a natural outcome of sound propagation. 
Barring reflection and dissipation by other means, all acoustic waves steepen into shocks in finite time \citep{1959flme.book.....L}.  

Shocks launched by waves from the convective zone have long been considered as a heat source for the solar corona \citep{1946NW.....33..118B, 1948ZA.....25..161B}. Yet, with few exceptions, existing solutions for shock formation and evolution from acoustic waves are restricted to relatively simple cases, such as planar and homogeneous or isothermal atmospheres.   We therefore seek more general solutions that can be applied to the stellar problems of interest, although we do consider only one-dimensional flow, such as spherical symmetry.

Spherical symmetry may appear to be a drastic simplification, as the Homunculus nebula, which surrounds the prototypical LBV $\eta$\,Car, is strongly bipolar.  Moreover, aspherical strong explosions are known to develop strongly non-radial flows near the stellar surface \citep{2013ApJ...779...60M,2014ApJ...790...71S}.

Nevertheless, a thorough understanding of the spherical problem is required for any detailed study of the non-spherical case, so this is where we begin.  The spherical idealization was also adopted in numerical investigations by \cite{2004MNRAS.354.1053W} and \cite{2010MNRAS.405.2113D}.
It allows us to describe the problem in simple terms: we start with a spherical hydrostatic stellar envelope of enclosed mass $m(r)$, density $\rho_0(r)$, pressure $p_0(r)$, and adiabatic sound speed $c_{s0}(r)$, where subscript `0' denotes undisturbed quantities, and consider the evolution of an outgoing sound pulse or wave train.   

Our ultimate goal is to predict (analytically, if possible) the entire sequence of events set in motion by a strong sound pulse from the stellar interior: its propagation as a sound wave; its steepening into a shock front; its strengthening into a strong shock, and arrival at the stellar surface; and the ensuing ejection and  fall-back of matter, and release of light.    

The strong-shock phase is pivotal to this sequence, because a normal strong shock must approach the stellar surface like the self-similar solutions identified by \cite{CPA:CPA3160130303}.   In these, the shock velocity  follows $v_s(r) \propto \rho_0(r)^{-\beta_1}$, for an eigenvalue $\beta_1\simeq 0.2$ that depends weakly on the density profile and the post-shock equation of state \citep{2013ApJ...773...79R}.    However, to calculate the coefficient to this strong-shock law, and to determine the pattern of shock-deposited heat and momentum, we must first analyze shock formation and strengthening. 

We focus here on the precise radius of shock formation at the end of purely acoustic propagation (Phase 1), and provide a simple estimate of the initial phase of shock strengthening in which the wave peak catches up with the shock front (Phase 2).  We postpone a detailed examination of shock evolution to a subsequent paper.  

We begin, therefore, by reviewing the nature of acoustic pulse propagation, before delving into a detailed analysis of weak shock formation and propagation.  We then derive a general expression for the condition of shock formation in two ways.  First, we use a wave action principle to generalize a classical derivation based on the crossing of sound front. Then, extending an analysis from the field of sonoluminescence, we use an expansion around the wave front to derive the same result.   Numerical simulations validate our result and provide insight into the subsequent phase of weak shock evolution.

\section{Propagation of a sound pulse} \label{sec:AcousticPropagation} 

Let us begin by considering the propagation of a sound pulse, launched outward from the stellar core into the stellar envelope.   Any such pulse can be decomposed into normal modes of the stellar envelope, and a pressure mode of angular momentum quantum number $\ell$ and frequency $\omega$ can only propagate through a zone with sound speed $c_{s0}$,  if $\omega^2 r^2 > \ell(\ell+1) c_{s0}^2$ \citep[e.g.,][]{1994sipp.book.....H}.    Non-radial modes ($\ell>0$) thus meet an angular momentum barrier and become evanescent inward of this radius.  This suppresses their generation by subsonic motions in the stellar core; we consider only radial motions, for which there is no such barrier.   

There is, nevertheless, an outer turning radius for radial sound waves.   Close to the stellar surface, where 
\begin{equation} \label{eq:TurningCriterion}
H \lesssim{ c_{s0}\over2\omega} \left(1 - 2{dH\over dr} \right)^{1/2},
\end{equation}  
the density scale height $H$ is traversed by sound in a time less than $\omega^{-1}$ and the atmosphere responds quasi-statically, causing reflection \citep[e.g.,][]{2010aste.book.....A}. (For later reference we designate  $\omega_{\rm ac}(r)$ as the local reflection frequency.) 

Away from its points of reflection, and in the absence of dissipation,  a linear, outwardly-propagating pressure wave carries a constant luminosity $L_w$.   It is worthwhile to understand why and when $L_w$ should be conserved, however; for this we rely on  \cite{1970PhFl...13.2710D}.

\citeauthor{1970PhFl...13.2710D} averaged the Lagrangian of an adiabatic fluid over a wave cycle, arriving at a conservation law $\partial n_w/\partial t + \nabla \cdot n_w {\mathbf v}_g =0$ for the wave action density \[n_w = {U_w\over \omega - {\mathbf k}\cdot{\mathbf u_0}}.\] Here ${\mathbf v}_g$ is the group velocity (${\mathbf v}_g = {\mathbf u}_0+ c_s \hat {\mathbf k}$ for sound waves), $U_w$ is the wave energy density, $\mathbf k$ is the wavevector, and $\mathbf u_0$ is the mean flow velocity.   In an otherwise motionless stellar envelope ($\mathbf u_0=0$), $U_w = \omega n_w$ is itself conserved, and if $\mathbf k$ is oriented radially outward, then the total wave luminosity \[ L_w = U_w c_{s0} A(r)\] across the area $A(r) = 4\pi r^2$ will be constant (along the wave trajectory) as the wave travels.    Different outgoing waves may nevertheless carry different values of $L_w$. 

Conservation of wave energy is a familiar feature of WKB theory.  Note, however, that the outward wave luminosity $L_w$ is {\em not} conserved if: (1) the wave is reflected;  (2) the stellar envelope is in motion, so that ${\mathbf k}\cdot{\mathbf u_0}$ varies;  (3) non-adiabatic effects lead to dissipation that saps the wave energy; or (4) a shock forms, as shocks involve localized dissipation.   

The mean wave energy density in a wave with peak velocity $u_w$ is $\bar U_w = \rho_0 u_w^2/2$, so the mean wave luminosity in a spherical star is 
\begin{equation} \label{eq:Lw_acoustic} 
\bar L_w = 4\pi r^2 \rho_0 c_{s0} {u_w^2\over 2} = {u_w^2\over c_{s0}^2} \Lwmax
\end{equation} 
where 
\begin{equation}\label{eq:Lw_max}
 \Lwmax(r)  \equiv \frac12 A(r) \rho_0 c_{s0}^3.
\end{equation}
We see immediately that $u_w(r)^2/c_{s0}(r)^2 = \bar L_w/\Lwmax(r)$.  Since supersonic wave motion ($u_w>c_{s0}$) produces a shock very rapidly, sound cannot propagate in zones where $\bar L_w > \Lwmax$ -- typically, the outer stellar envelope or atmosphere.   However, wave dissipation by diffusion or shock formation sets in far before this condition is satisfied.   A shock-driven outburst is only possible if diffusion does not sap $L_w$ prior to shock formation, so it is important to examine both processes in detail. 

\subsection{Thermal diffusion} \label{SS:diffusion} 

Our estimate of losses due to thermal diffusion will be approximate, and similar to the analysis by \citet{2012MNRAS.423L..92Q}.  Below the stellar photosphere, this process  is described by the diffusion equation $F_{\rm rad} = - \nu \nabla U_{\rm rad}$, where $F_{\rm rad}$ is the diffusive flux, $\nu$ is the thermal diffusivity, and $U_{\rm rad}$ is the portion of the total energy density $U_{\rm th}$ that can diffuse (i.e., the radiation energy density, if diffusion is due to photons).   This equation applies to the outward diffusion of luminosity, so 
\[ \nu = {L_{\rm rad}\over4 \pi r^2 |\nabla U_{\rm rad}|} = {L_{\rm rad} \Hprad \over 4\pi r^2 U_{\rm rad}}.\] 
Here $\Hprad = U_{\rm rad}/|\nabla U_{\rm rad}|$ is the radiation pressure scale height, and $L_{\rm rad}$ is the diffusive portion of the stellar luminosity $L$, so $L_{\rm rad}(r)\leq L(r)$, where the equality holds in regions that are not convective. 

If we consider the change of the wave luminosity $L_w(\varphi)$ along a wave front (phase $\varphi=$const.) due to thermal diffusion, then considering that the thermal diffusion time is $c_{s0}^2/(\omega^2 \nu)$, we find \[ \dot L_w(\varphi) = c_{s0} {dL_w(\varphi)\over dr}  \simeq - {L_{w,{\rm rad}}(\varphi)  \nu \omega^2\over c_{s0}^2}.\] Here $L_{w,{\rm rad}}\simeq L_w U_{\rm rad}/U_{\rm th}$ is the part of the wave luminosity subject to diffusion.    Using our expression for $\nu$, the net loss across a distance $\Hprad$ is 
\begin{eqnarray} \label{eq:DiffusiveLoss_General} 
\Hprad   { |dL_w(\varphi)/dr| \over L_w(\varphi)} &\simeq& {\omega^2 \Hprad^2 L_{\rm rad} \over 4\pi c_{s0}^3 r^2 U_{\rm th} }  \\ &=& {1\over 8\gamma(\gamma-1) }\left(2  \omega \Hprad \over c_{s0}\right)^2 {L_{\rm rad}\over\Lwmax}. \nonumber
\end{eqnarray} 
The last step relies on the relation $U_{\rm th} = \gamma(\gamma-1)\rho_0 c_s^2$ for an ideal fluid of adiabatic index $\gamma$, and yields a numerical prefactor in the range 0.1 to 0.3.  

The form of (\ref{eq:DiffusiveLoss_General}) is convenient for determining whether linear acoustic waves will damp or reflect as they approach the stellar surface.  However, our purpose is to quantify the non-linear process of shock formation, which we consider next.  We return to radiative damping in section \ref{S:ShocksVersusDiffusion}, where we derive a critical wave luminosity for shock formation. 

\section{Shock Formation}\label{sec:stages}
The evolution of an acoustic wave into a shock has two distinct stages: the creation of the shock discontinuity, and the driving of this shock by the acoustic pulse behind it.  If this driving is successful, the shock will become strong and approach the stellar surface in the manner described by \citet{CPA:CPA3160130303}.  We seek to predict the wave evolution through the first and second stages.  We first identify the radius $\Rsf$ at which the shock first forms; for this, we provide both a heuristic and a detailed calculation.

\subsection{Shock formation radius: heuristic derivation} \label{SS:Heuristic_shock_formation} 

 For a heuristic derivation of the location of shock formation, consider the fact that acoustical information (such as the value of $L_w$) travels along outward-moving sound fronts (or characteristics), and that a shock forms when characteristics arrive at the same location carrying conflicting information.  Some variation of the propagation speed $u+c_s$ is inevitable if $u$ is non-uniform, as the scalings of adiabatic linear perturbations \citep{1959flme.book.....L,whitham1974linear} imply 
 \begin{equation}\label{eq:deltaUplusC} \delta(u + c_s)={ \gamma+1\over2}\delta u. \end{equation} 
 Here $\delta$ represents the perturbation from the background state for a given fluid element.   In this section we generalize a classic result \citep[][\S 101]{1959flme.book.....L} to non-planar and non-uniform environments.  
 
Consider two nearby characteristics launched from an initial radius $r_i$ but separated in space by a small amount $\Delta r_i$ at an initial time $t_i$ -- or equivalently, separated in time (at a fixed $r_i$) by $\Delta t_i = -\Delta r_i/c_{s0}(r_i)$.    Each propagates outward at $dr= (u+c_s)dt$, but they move at different speeds because $L_w$ differs slightly between them.   At some larger radius, the difference in arrival times is 
 \[
 \Delta t = \Delta t_i + \Delta \int_{r_i}^r {dr'\over u + c_s } \simeq  \Delta t_i  - \Delta \int_{r_i}^r {dr'\over c_{s0}} {\gamma+1\over 2} { \delta u \over c_{s0}}.  
 \] 
 The second step uses equation (\ref{eq:deltaUplusC}), as well as $(u + c_s)^{-1} = [c_{s0} + \delta(u+c_s)]^{-1} \simeq c_{s0}^{-1} [1- \delta(u+c_s)/c_{s0}]$, which is correct to first order in $\delta(u+c_s)/c_{s0}$.    
 
 We can now employ the conservation of wave luminosity, in the form \[{u\over u(r_i)} ={c_{s0}\over c_{s0}(r_i) } \sqrt{\Lwmax(r_i)\over \Lwmax}.\]   If we also write $\Delta u(r_i) = (\partial u/\partial r)_i \Delta r_i$, then after a little algebra we arrive at the shock formation condition 
 \begin{equation} \label{eq_HeuristicShockFormationCriterion}
 \left(-{\partial u\over \partial r}\right)_i \int_{r_i}^{\Rsf} {\gamma(r)+1\over 2}\sqrt{\Lwmax(r_i)\over \Lwmax(r) }{dr\over c_{s0}(r)} = 1
 \end{equation} 
 which corresponds to the crossing of characteristics: $\Delta t=0$ at $r=\Rsf$. 
 
This is only a heuristic derivation, as we relied on \citeauthor{1970PhFl...13.2710D}'s phase-averaged conservation law to infer the constancy of $L_w$, and from that, the propagation of characteristics within a single wave pulse.   This procedure is hardly rigorous.   However, we now show that equation (\ref{eq_HeuristicShockFormationCriterion}) coincides perfectly with a more detailed calculation.

\begin{figure*}
	\centering
    \includegraphics[width=\textwidth]{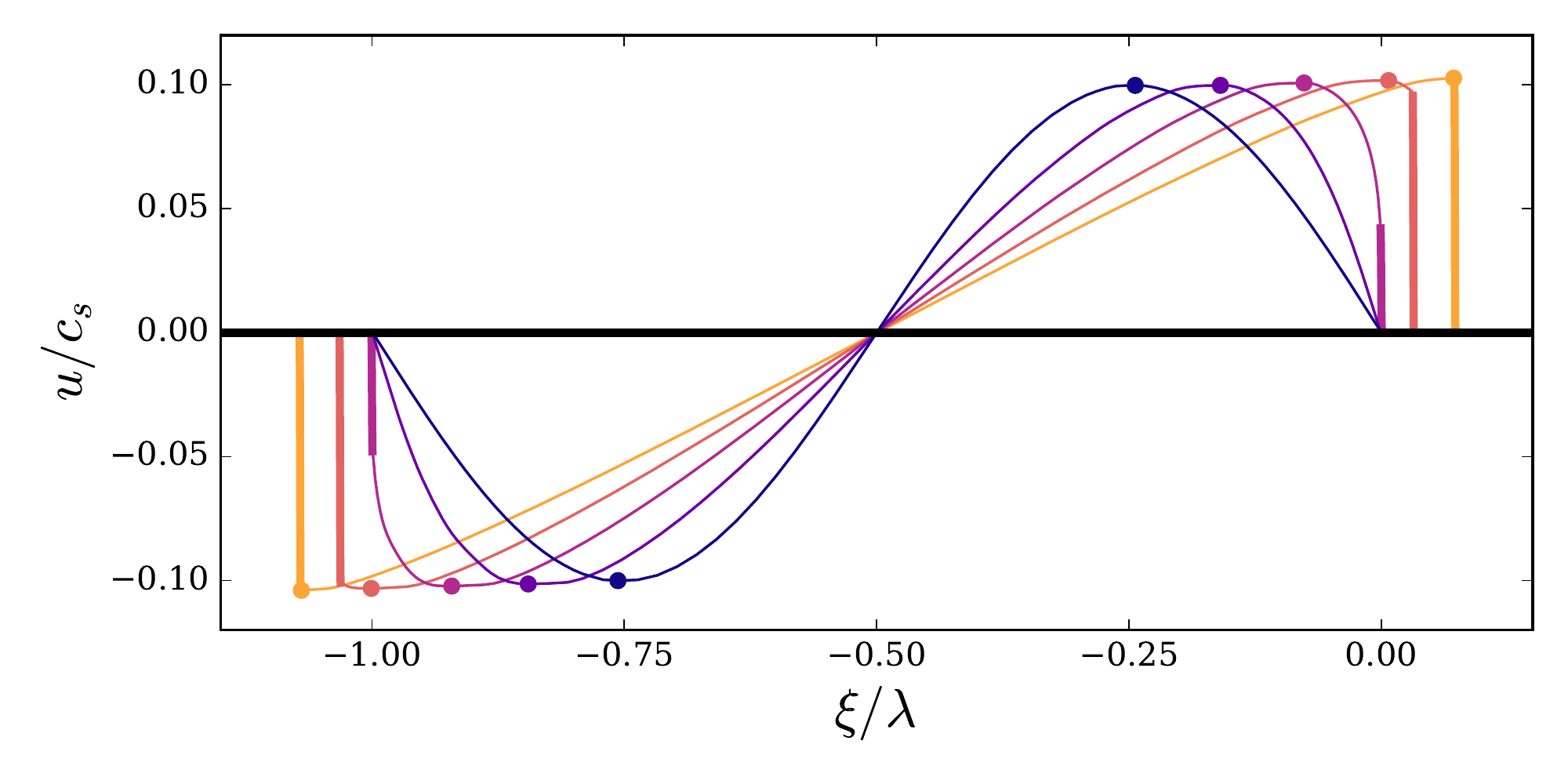}
    \caption{The deformation of a single sinusoidal impulse in a planar, isothermal atmosphere. The wavelength remains constant until a shock forms in the third frame. This indicates the end of stage one and beginning of stage two. The shockwave becomes fully developed and stage two completes once the wave peak (point) coincides with the shock location.  }
    \label{fig:waveform}
\end{figure*}

\subsection{Detailed derivation of shock formation radius}\label{SS:ls01}

The condition for shock formation from a sound pulse has been worked out in the context of sonoluminescence by \citeauthor{FLM:68109} (\citeyear{FLM:68109}, hereafter LS01), and the solution is applicable to inertially-confined fusion and related topics.  We generalize LS01's analysis to account for a body force (due to the stellar gravity $g$) as well as the variations of fluid properties that define the stellar structure.  

We begin with the Euler equations,
\bea
\pt_t \rho + u\pt_r \rho + \rho\pt_r u +\frac{\alpha \rho u  }{r}&=& 0, \label{e_plane_1} \\
\pt_t u + u\pt_r u + \frac{1}{\rho}\pt_r p &=& -g, \label{e_plane_2} \\
\pt_t p + u\pt_r p + \rho c_s^2\pt_r u &=& 0,\label{e_plane_3}
\eea
where the density $\rho$, pressure $p$, sound speed $c_s$, and fluid velocity $u$ reference the wave properties, and $\pt_r$ and $\pt_t$ are partial derivatives with respect to space $r$ and time $t$.  In addition to the spherical case ($\alpha=2$) we allow for cylindrical and planar cases ($\alpha = 1$ and 0, respectively); in general $A(r) = 2\pi^{(\alpha+1)/2} r^{\alpha}/\Gamma[(\alpha+1)/2]$.

We assume the structure of the quiescent gas is known (i.e., $\rho_0(r)$, $p_0(r)$, $c_{s0}(r)$) and permit the quiescent adiabatic index,
\beq
\gamma_0 \equiv \frac{d \mrm{ln}(\rho_0 c_{s0}^2)}{d \mrm{ln}(\rho_0)},  \label{e_gamma_0} 
\eeq
to vary across the star: $\gamma_0=\gamma_0(r)$. This allows for a non-uniform composition in gas and radiation.  We  account for variations in the instantaneous adiabatic index 
\beq
\gamma \equiv \frac{d \mrm{ln}(p)}{d \mrm{ln}(\rho)} = \frac{\rho c_s^2}{p}  \label{e_gamma}
\eeq
which differs from $\gamma_0$ as a fluid element is perturbed. We neglect effects from ionization, which may absorb heat. 

Taking equations (\ref{e_gamma_0}) and (\ref{e_gamma}) with the Euler equations (\ref{e_plane_1})-(\ref{e_plane_3}), we choose to substitute $\rho$ with $c_s$ to work only with variables $p$, $u$, $c_s$, and $\gamma$. This generates the following differential equations:
\bea
\pt_t c_s + u\pt_r c_s + q c_s\pt_r u + \frac{\alpha q c_s u}{r} &=&  \frac{c_s}{2\gamma}\frac{d \gamma}{dt}, \label{e_diff_1}\\
\pt_t u +  u\pt_r u + \frac{c_s^2}{\gamma p}\pt_r p &=&-g, \label{e_diff_2}\\
\pt_t p + u\pt_r p + \gamma p \pt_r u &=& -\frac{\alpha\gamma p u}{r},\label{e_diff_3}
\eea
where
\beq
\frac{d \gamma}{d t} \equiv \pt_t \gamma + u\pt_r\gamma = (\gamma-\gamma_0)\pt_r u, 
\label{e_diff_4}
\eeq
and $q\equiv (\gamma-1)/2$. We assume the body force is independent of perturbations (Cowling's approximation). The wave travels only in the radial direction, so refraction is ignored.

\cite{whitham1974linear} found that a Taylor expansion of fluid properties about the wave front generates a closed system of equations. From these equations, \citeauthor{FLM:68109} found velocity gradient $\partial_ru(r)$ evolution can be described in an explicit and analytic form until shock formation $\partial_ru\rightarrow-\infty$. While much of the following derivation is similar to LS01, we include a body force (eg. gravity), cylindrical wave solutions, and a variable adiabatic index.

The wave front $r = F(t)$ propagates outward (or left to right) at the local quiescent sound speed,
\begin{equation}
F'(t)=c_{s0}(F(t)).
\label{e_wf_speed}
\end{equation} 
The primes are derivatives with respect to their independent variable. We define a new coordinate variable $\xi=r-F(t)$ around the wave front ($\xi = 0$) and expand the fluid variables for $\xi<0$:
\bea
c_s(\xi,t) &=& c_{s0}(F(t)) + \xi c_{s1}(t) + \frac{1}{2}\xi^2c_{s2}(t) +... \ ,\\
u(\xi,t) &=& \xi u_1(t) + \frac{1}{2}\xi^2u_2(t) + ... \ ,\\
p(\xi,t) &=& p_0(F(t)) + \xi p_1(t) + \frac{1}{2}\xi^2p_2(t)+... \ , \\
\gamma(\xi,t) &=&  \gamma_0(F(t)) + \xi \gamma_1(t) + \frac{1}{2}\xi^2 \gamma_2(t) +...   \label{e_gamma_expand} 
\eea
Integer subscripts represent the number of spatial derivatives taken (e.g., $c_{si}=(\partial_r)^ic_s$). Since the wave front is also a node, we use the quiescent gas values for variables with subscript 0. Variables with non-zero subscript are spatial gradients evaluated at the wave front and are functions in only time. Therefore, our notation states $u_1'(t) = du_1/dt$ and $p_0' = dp_0/dr$. Note that we assume the quiescent gas is initially static $u_0=0$. 

Next, we substitute the expanded variables into the set of differential equations (\ref{e_diff_1})-(\ref{e_diff_4}). Since the derivatives are with respect to $r$ and not $\xi$, we change the variables to generate a new derivative:
\bea
\left[\pt_t\right]_r = \left[\pt_t\right]_{\xi}+ \left[\pt_t (\xi)\right]_r \pt_{\xi} &=& \pt_t - F'(t)\pt_{\xi} \nonumber \\
&=& \pt_t - c_{s0}(F(t))\pt_{\xi}.
\eea

We collect the $\xi^0$ and $\xi^1$ terms from each differential equation to obtain eight equations, which are listed in the Appendix. This requires a meticulous account of all variables. Combining these equations together presents a single differential equation about the variable $u_1$, which measures the wave steepness or gradient, as a function of only the quiescent gas,
\bea
0 &=& 2u_1' + \left(\gamma_0 + 1 \right)u_1^2  \nonumber \\
& \ & \ \ \ \ \ \ + \left(c_{s0}' + \frac{\gamma_0'}{\gamma_0}c_{s0} + \frac{\alpha c_{s0}}{r} - \gamma_0\frac{g}{c_{s0}}\right)u_1.
\eea
This is an example of a Bernoulli equation \citep{ince1956ordinary}, which has an analytic solution of the form 
\bea
u_1^{-1}(r) &=& \sqrt{\frac{\Lwmax(r)}{\Lwmax(r_i) } }\times   \label{e_wavefront} \\
&\ & \left[ u_1^{-1}(r_i) +  \int_{r_i}^r  \frac{\gamma_0(r')+1 }{2} \sqrt{\frac{\Lwmax(r_i) }{\Lwmax(r')  } }\frac{dr'}{c_{s0}(r')}  \right].   \nonumber\ \
\eea

Although shock formation is a nonlinear process, our Taylor expansion is justified by the fact that the shock forms at the wave node; only the first term $u_1$ appears in this solution.  Insofar as the combination $c_{s0}^{-1} \Lwmax^{-1/2} r$ tends to be much larger where a wave shocks than where it was launched (at least in the stellar context), it is reasonably accurate to evaluate equation (\ref{e_wavefront}) with $r_i\rightarrow0$. 

Comparison to equation (\ref{eq_HeuristicShockFormationCriterion}) shows that shock formation ($u_1\rightarrow \infty$) occurs precisely where our heuristic analysis predicts (i.e., $r=\Rsf$).   The wave front evolution is defined entirely by the structure of the quiescent gas, initial wave front gradient 
$ u_1(r_i)$ and location $r_i = F(0)$, and {\em not} on the wave's other properties (wavelength, amplitude, etc.).   And, our result holds equally well for planar, cylindrical, or spherical symmetry,    and for inward as well as outward propagation.  Shock formation is used in fields as diverse as the heating of the Solar corona \citep{1961ApJ...134..347O}, the deflagration-to-detonation transition in type Ia supernovae \citep{2013A&A...550A.105C}, and sonoluminescence (LS01), among others, so this general result should be widely applicable. 

 Although shock formation is a purely local process on the most rapidly compressive characteristic, we can relate it to the properties of a larger wave or pulse. Suppose the wave is monochromatic with initial peak velocity amplitude $u_w(r_i)$. The peak compression rate is $\max[-u_1(r_i)] =  \omega u_w(r_i)/c_{s0}(r_i)$, achieved at the wave node.  In our shock formation criterion, the combination $\max[-u_1(r_i)] \sqrt{\Lwmax(r_i)}$ is equivalent to $\bar L_w(r_i)$; but this equals $\bar L_w$ elsewhere, so long as the conditions discussed at the start of \S\ref{sec:AcousticPropagation} hold.   Condition (\ref{eq_HeuristicShockFormationCriterion}) therefore becomes 
\beq\label{eq:whereWaveShocks}
\int_{t_i}^{t_{\rm sf}}  {\gamma_0+ 1 \over 2} \sqrt{\bar L_w\over \Lwmax} \omega\, dt =1; 
\eeq
 $\omega\, dt$ is the change of phase angle.  In other words, the wave propagates for \[\left<{\gamma_0+1\over 2}\sqrt{\bar L_w\over \Lwmax}\right>^{-1}\] radians before producing a shock, where the bracket means a time average along the wave front.   The total propagation time is proportional to $\bar L_w^{-1/2} \omega^{-1}$, so stronger and higher-frequency waves shock earlier.

\subsection{Maturation of the shockwave}

To estimate the point of intersection between the wave peak and shock front, we derive their respective trajectories. First, suppose a monochromatic wave with luminosity $\bar L_w$ and frequency $\omega$ is led by a compressive edge (i.e., $u_1(r_i)<0$). The wave peak initially lags behind the wave front by a distance $ r_i-r_w = \lambda/4 = \pi c_{s0}/(2\omega)$. 

The peak propagates with a speed $v_w=u_w+c_{s,w}$, where $c_{s,w}$ is the compressed local sound speed. Assuming properties of the gas do not vary significantly under compression (i.e., constant $p/\rho^{\gamma_0}$ and $\gamma=\gamma_0$), we can write the compressed sound speed $c_{s,w}^2=\gamma_0(p_0/\rho_0^{\gamma_0})^{1/\gamma_0}p_w^{(\gamma_0-1)/\gamma_0}$ in terms of the peak pressure $p_w$. The thermodynamic expression of the mean wave energy density $\bar U_w=(p_w-p_0)^2/2\rho_0c_{s0}^2=\bar L_w/4\pi r^2 c_{s0}$ allows $p_w$ and $c_{s,w}$ to be expressed in terms of conserved wave properties,
\beq
c_{s,w}^2=c_{s0}^2  \left(1 + \gamma_0\sqrt{\frac{\bar L_w}{L_{\rm max} }} \right)^{\frac{\gamma_0-1}{\gamma_0}}.\nonumber
\eeq
With the kinematic expression (\ref{eq:Lw_acoustic}), the speed of the wave peak becomes
\begin{equation}
    \frac{v_w}{c_{s0}} = \sqrt{\frac{\bar L_w}{L_{\rm max}}} + \left(1 + \gamma_0\sqrt{\frac{\bar L_w}{L_{\rm max} }} \right)^{\frac{\gamma_0-1}{2\gamma_0}}.
\end{equation}

Before shock formation ($r<R_{\rm sf}$), the wave front simply travels at the local sound speed $c_{s0}$. For $r>R_{\rm sf}$, the exact shock velocity requires numerical calculations to describe the arrival of the remaining wave pulse. We circumvent this calculation by approximating the shock front strength $z=p_s/p_0$ with the wave peak properties ($z\simeq p_w/p_0$). With the following shock jump condition, 
\begin{equation}
    z =  \frac{2\gamma_0 M_s^2 - (\gamma_0-1)}{\gamma_0+1},
    \label{e_shock_jump}
\end{equation}
we approximate the shock Mach number $M_s = v_s/c_{s0}$ to be
\begin{equation}
    M_s^2 \simeq 1+\frac{\gamma+1}{2}\sqrt{\frac{\bar L_w}{L_{\rm max}}}.
    \label{e_mach}
\end{equation}

Thus, the wave peak and shock front converge at the same location $r=R_s$, once
\begin{equation}
    \int_{r_i - \lambda/4}^{R_s} \frac{dr}{v_p}= \int_{r_0}^{R_{\rm sf}} \frac{dr}{c_{s0}} + \int_{R_{\rm sf}}^{R_s} \frac{dr}{v_s}
    \label{e_full_shock}
\end{equation}
is satisfied.

\section{Hydrodynamic Simulations}\label{sec:flash}
To test our analytical predictions of shock formation and provide examples of shock evolution and strengthening, we turn to numerical simulations. We construct  one-dimensional planar and spherical simulations in  FLASH \citep{flash00}, a hydrodynamic adaptive mesh refinement (AMR) code. All quiescent structures begin in hydrostatic equilibrium with a gravitational acceleration $g(r)$ that is independent of fluid perturbations (Cowling's approximation). All simulations have inner reflecting and outer diode (outflow) boundary conditions. Two structures are considered: a planar isothermal atmosphere and a $n=3$ stellar polytrope. The adiabatic index $\gamma_0$ is held fixed in all of our simulations. 

\subsection{Planar Earth Atmosphere} \label{sec:planar}
For planar shock formation we consider a vertically stratified, initially isothermal atmosphere of gas (FLASH's default Earth atmosphere) with $\gamma_0=1.4$, in constant gravity.  Lengths in this section should be compared to the 8.8\,km scale height of the model, although the results can be scaled to any similar atmopshere.

Upward-travelling waves are initialized by setting the isothermal atmosphere out of equilibrium; 
an example of the initial waveform and its evolution is shown in Figure \ref{fig:waveform}. The initially sinusoidal wave steepens (Phase 1) and forms a shock at the wave node, after which the wave peak approaches (Phase 2) and merges with the shock front.   

\begin{figure}
	\centering
    \includegraphics[width=\columnwidth]{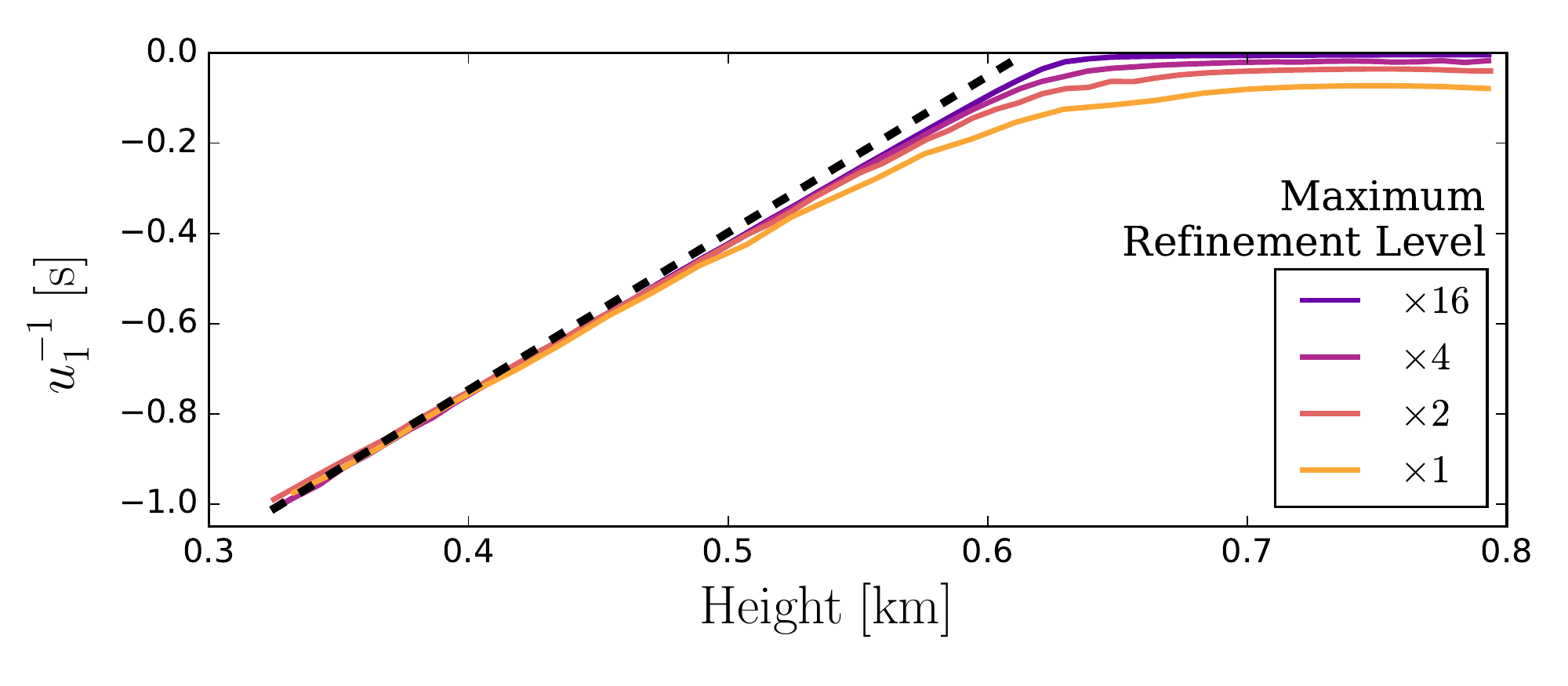}
    \caption{Evolution of the wave front gradient $u_1=\partial_r u $ of four waves with different grid resolutions. The wave may manifest manifest across $\sim3200$ (AMR) grid cells for the finest resolution ${\rm min}(\Delta x) \gtrsim \lambda / (200 \times [1, 2, 4, 16])$.  All initial wave properties are A2 from Table \ref{tab:planar}. The dashed line is the analytic wave steepening prediction (eq. \ref{e_wavefront}).  }
    \label{fig:resolution}
\end{figure}

In a grid-based simulation, a shock must span multiple grid cells of length $\Delta x$; for a given velocity jump $\Delta u$ this sets an upper limit to the compression rate $-u_1$ of order  $\Delta u / \Delta x$.  Given this limitation, we expect the numerical solution to converge toward the analytical prediction (eq.\ \ref{eq:whereWaveShocks} as $\Delta x\rightarrow 0$; this verified in Figure \ref{fig:resolution} for one set of initial conditions.

\begin{figure}
	\centering
    \includegraphics[width=\columnwidth]{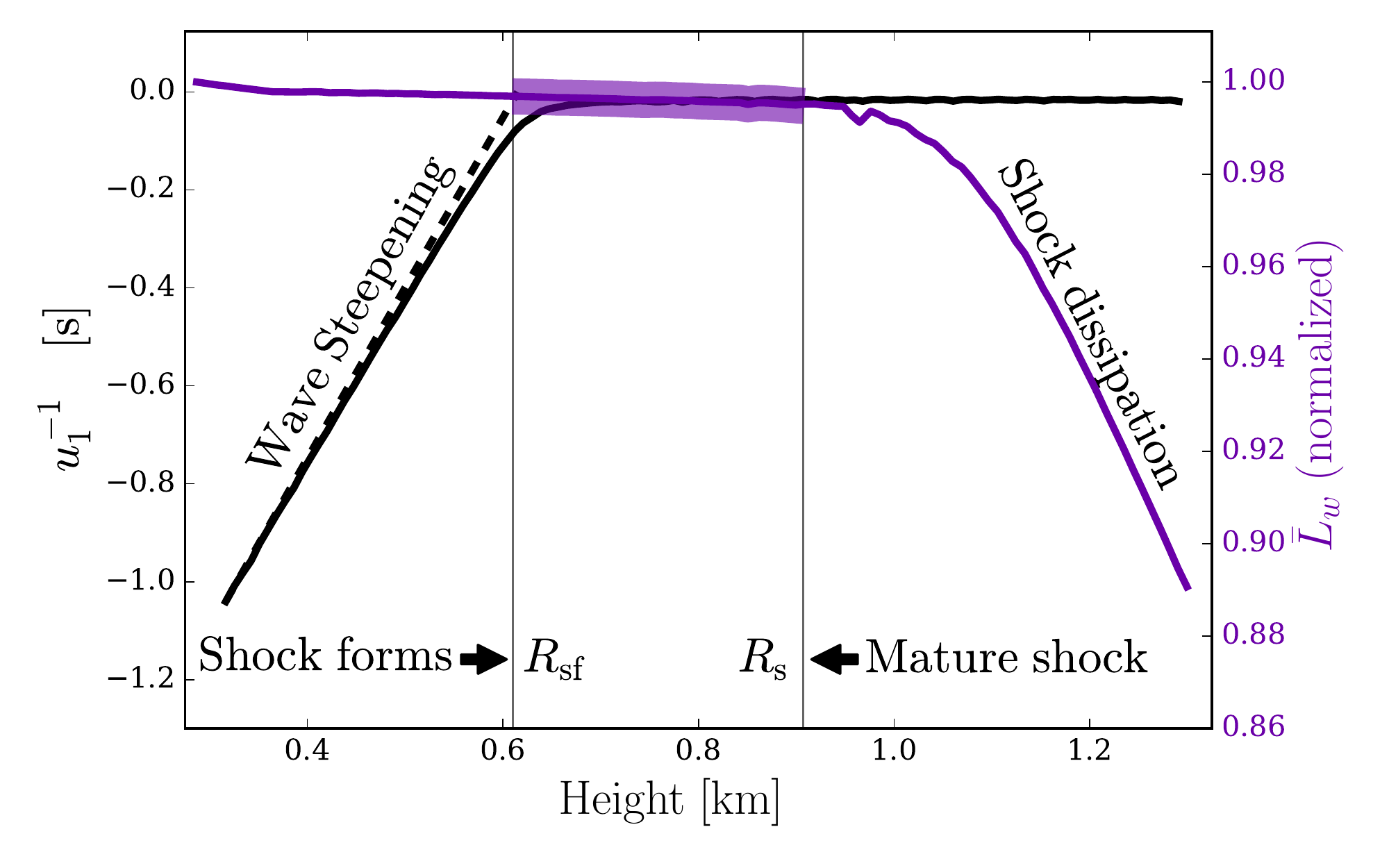}
    \caption{Stages and evolution of the wave front gradient (solid black) and peak wave luminosity (purple). The thick coloured band indicates our predictions for Phase 2, which begins with shock formation and ends when the wave peak reaches the shock front. 
    Once a shock fully develops, the peak acoustic wave luminosity $L_w=2\pi r^2 (p_w - p_0)^2/(\rho_0 c_{s0})$ declines due to shock dissipation.  }
    \label{fig:grand_planar}
\end{figure}

Figure \ref{fig:grand_planar} shows that the peak wave luminosity is conserved during the acoustical phase of propagation (Phase 1), just as predicted in \citeauthor{1970PhFl...13.2710D}'s theory. Moreover it is diminished by less than 0.5\% in the time between shock formation and the arrival of the wave peak at the shock front (Phase 2), and numerical dissipation is responsible for part of this loss.

\begin{table}
  \caption{Wave characteristics, planar simulations.}
  \label{tab:planar}
  \begin{center}
    \leavevmode
    \begin{tabular}{crcc} \hline \hline              
     Label & $\lambda$ (m)    & $u_p/a_0$    &  $\lambda a_0/u_p$               \\ \hline 
A1 &  50 & 0.02    & 2500           \\
A2 &  250 & 0.10  & 2500 \\
B1 & 50 & 0.04  &   1250       \\
B2 &  250 & 0.20    &1250       \\ \hline
    \end{tabular}
  \end{center}
  \tablecomments{The atmospheric scale height is 8777\,m and the base resolution corresponds to $\Delta x=1.25$\,m. }
\end{table}

\begin{figure}
  \centering
\begingroup
\captionsetup[subfigure]{width=1\columnwidth}
 \subfloat[Wave front gradient] {\label{fig:planar_sf}\includegraphics[width=\columnwidth]{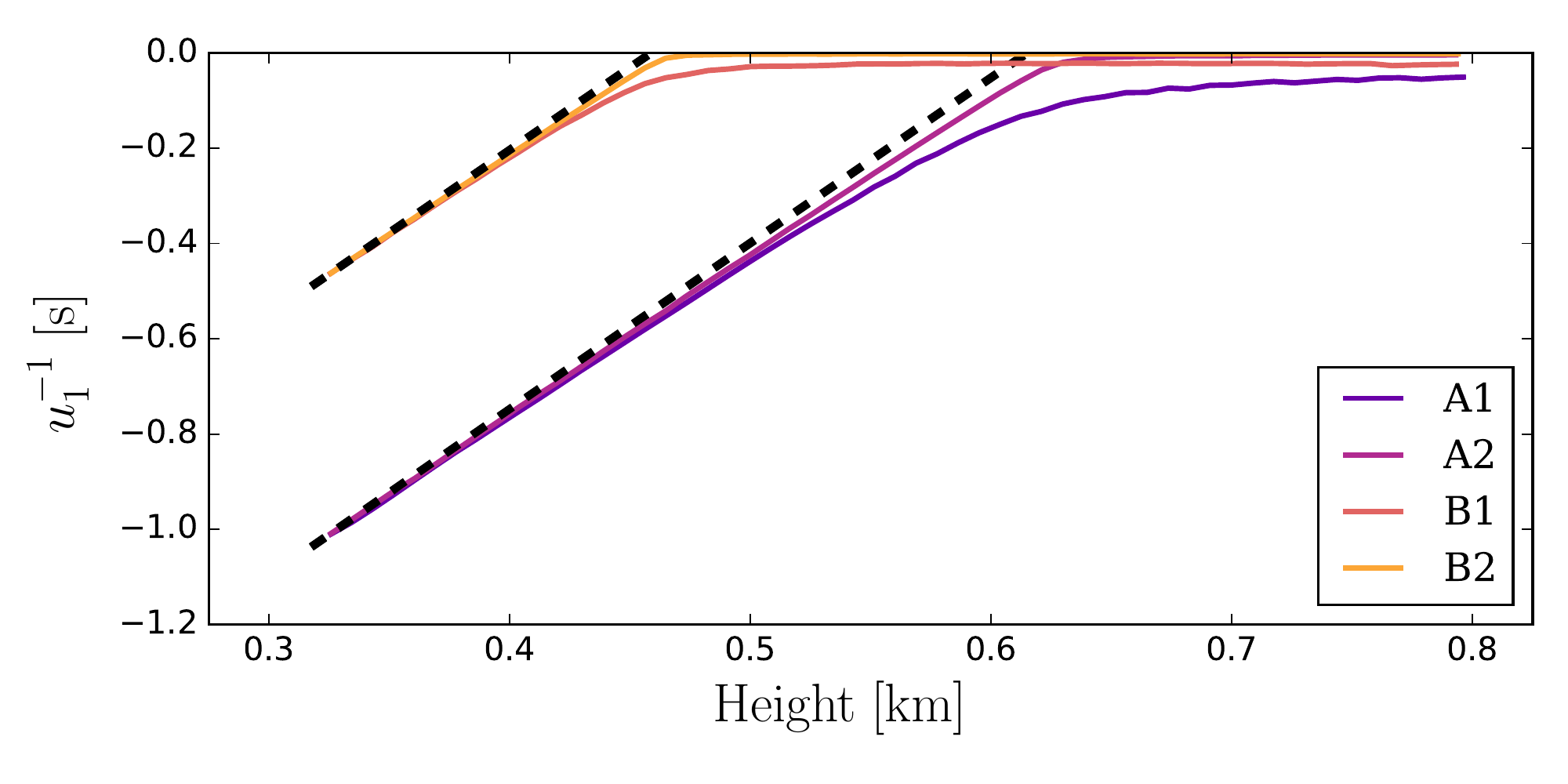}}\\
  \subfloat[Peak wave luminosity ] {\label{fig:planar_sd}\includegraphics[width=\columnwidth]{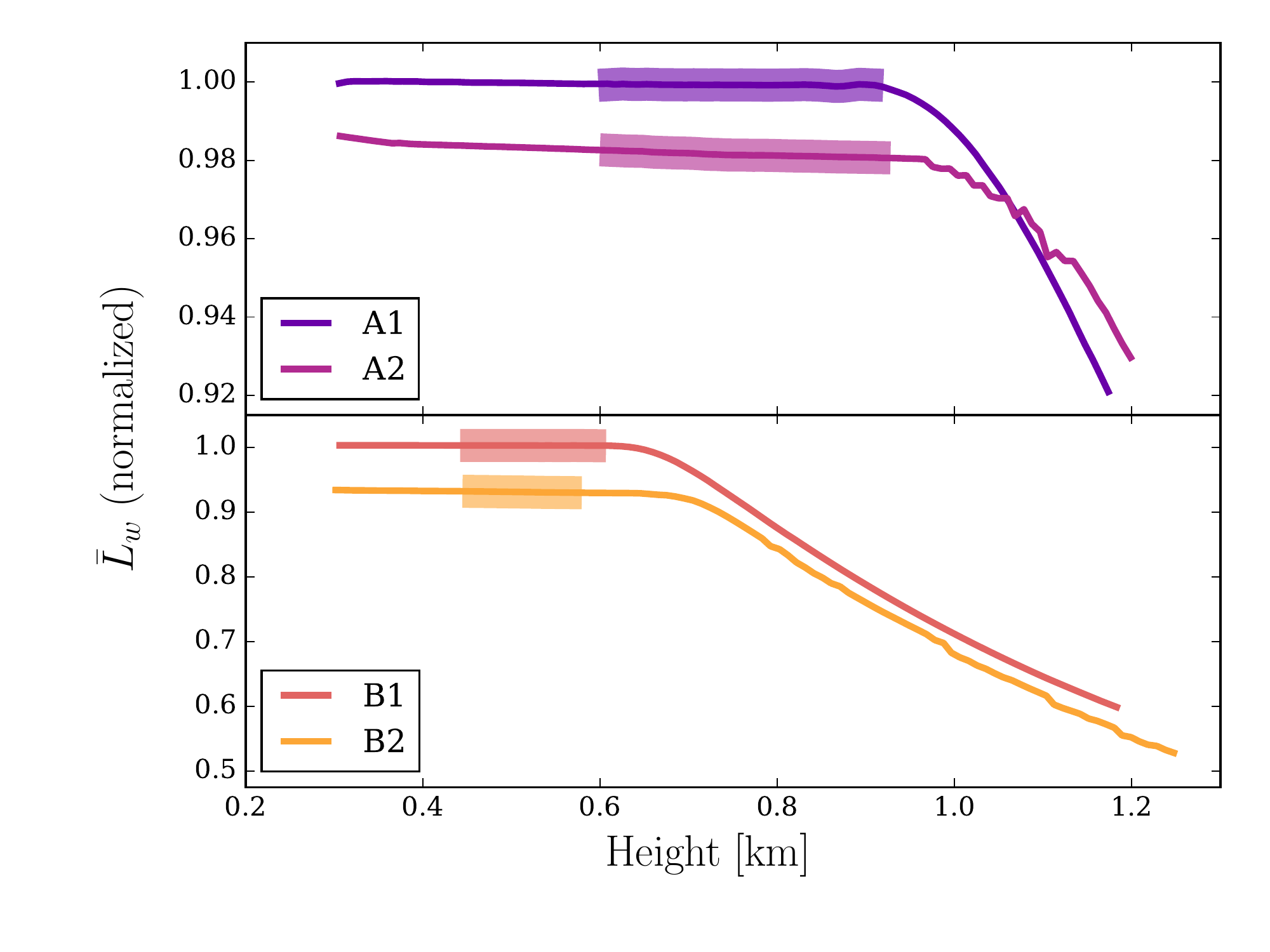}}\\

\endgroup
\caption[]{Numerical shock formation from vertically-propagating waves in a planar isothermal atmosphere (scale height 8.8\,km). Waves within sets A and B have the same initial wave front gradient and different wavelengths and amplitudes (see Table \ref{tab:planar}). See Fig. \ref{fig:resolution} and \ref{fig:grand_planar} for figure descriptions.  }
\label{fig:planar_simulation}
\end{figure}

We launch four waves as described in Table \ref{tab:planar}.  Within groups A and B, waves have identical $u_p/\lambda$ and maximum compression rate. Equation (\ref{eq:whereWaveShocks}) therefore states that waves in each group will steepen identically and form shocks $u_1^{-1}=0$ at the same location; this is confirmed in Figure \ref{fig:planar_sf}. A1 and B1 have shorter wavelengths and are more poorly resolved, so they obey the analytical prediction more poorly than A2 and B2. 

\subsection{Spherical Polytropes}
We interpolate a $n=3$ polytropic stellar model with a constant adiabatic index of $\gamma_0=4/3$ onto a uniform grid of 130,000 cells.  We disable AMR, as mesh refinement appeared to stimulate spurious oscillations in regions with short scale heights. We nevertheless observe small oscillations and a weak outflow of matter and corresponding inward-moving rarefaction wave due to imperfect force balance and the outer boundary conditions.  (While density and pressure are formally zero at the stellar surface, simulation fluid variables cannot be defined zero. As a result, the grid boundary lies inside the stellar radius.)  However, these have negligible impact on our results.  

Our initial conditions generated both inward and outward travelling waves, so we measure the outgoing wave properties after it has separated from the ingoing wave. Waves with the shortest wavelength initially span 1\% of the domain, contracting to 0.5\% of the domain as they traverse regions of lower sound speed.  However this is still highly resolved (650 cells).  The wave luminosity is conserved to within 1\%. 

\begin{figure}
  \centering
\begingroup
\captionsetup[subfigure]{width=1\columnwidth}
 \subfloat[Wave front gradient] {\label{fig:sph_sf}\includegraphics[width=\columnwidth]{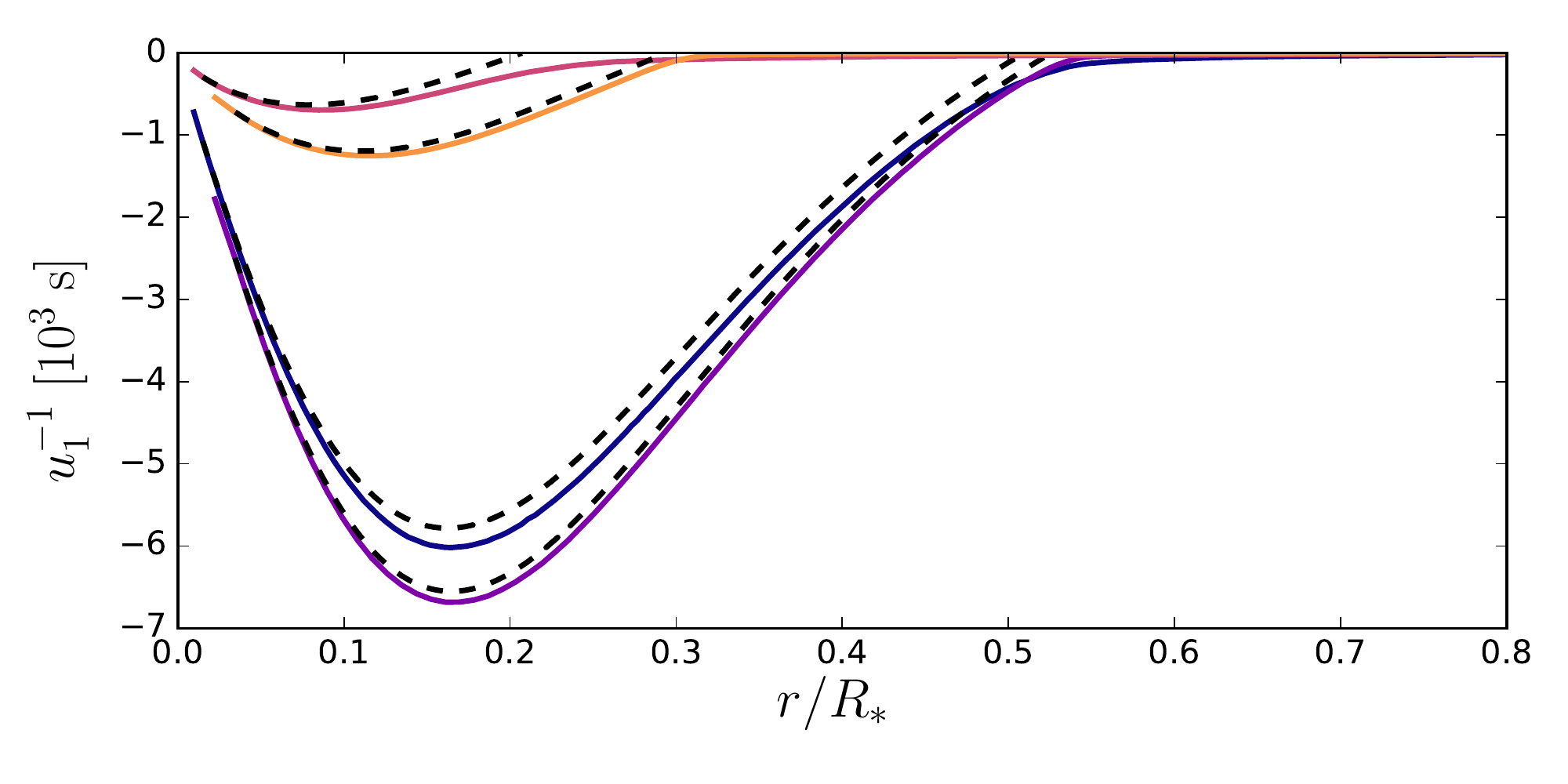}}\\
  \subfloat[Peak wave luminosity ] {\label{fig:sph_sd}\includegraphics[width=\columnwidth]{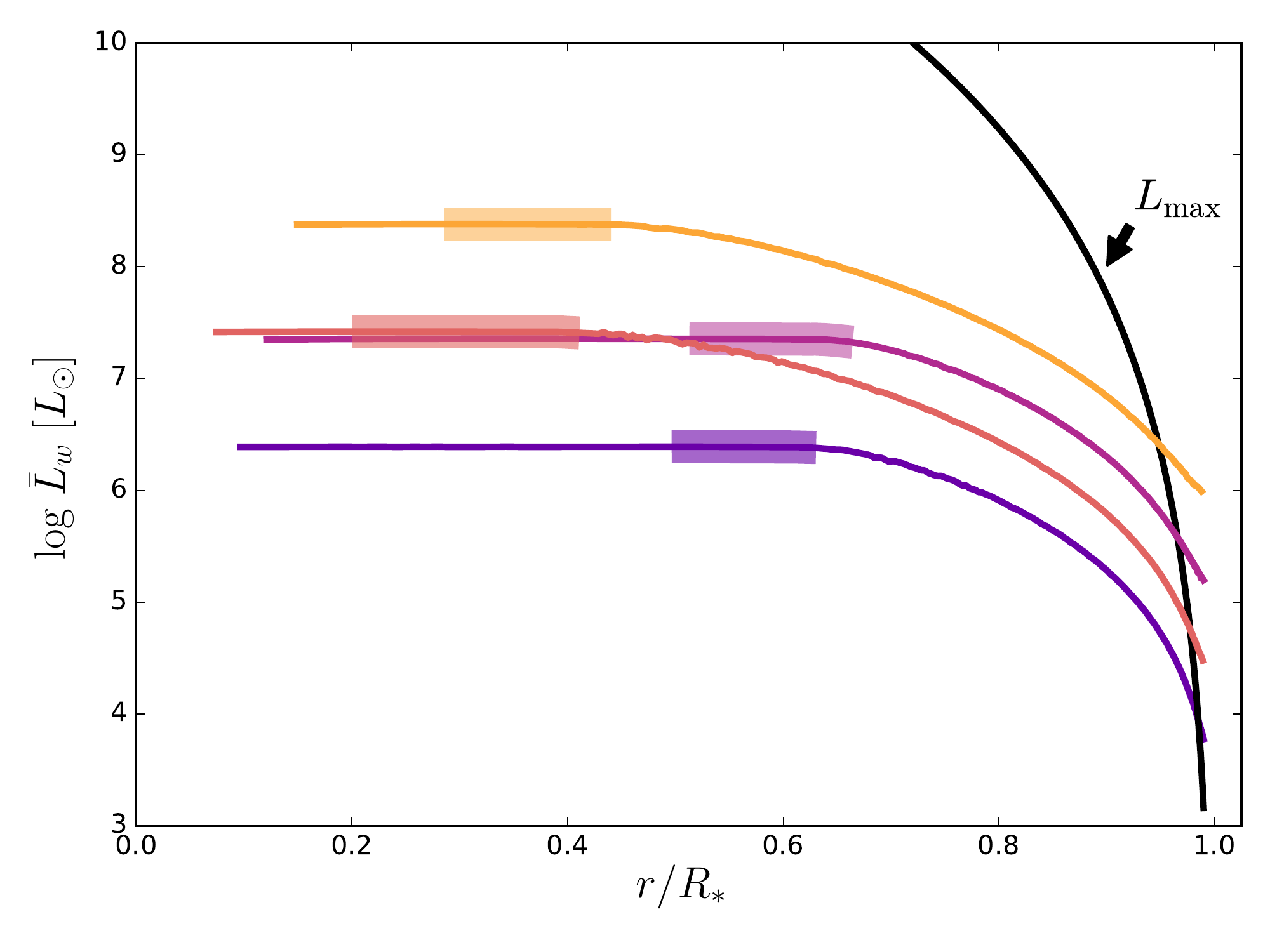}}\\

\endgroup
\caption[]{Numerical results from the launching of four waves of various strengths and frequencies in a $n=3$ stellar polytrope.

See Fig. \ref{fig:resolution} and \ref{fig:grand_planar} for figure descriptions. A shock becomes strong where the post-shock wave luminosity $L_w$ exceeds the maximum acoustic luminosity $\Lwmax$ (black line). 
}
\label{fig:spherical_simulation}
\end{figure}

From Figure \ref{fig:sph_sf}, we observe equation (\ref{eq:whereWaveShocks}) to successfully predict the location of shock formation for a $n=3$ stellar polytrope in all of our simulations. Our estimate for where the shock fully develops $R_s$ is accurate to within $|R_s - R_{s,{\rm sim}}|=1-2$ local wavelengths or $\lesssim 6\times10^{-3}\ R_*$.

\section{Shock dissipation or radiative damping?} \label{S:ShocksVersusDiffusion}

The results of the previous sections allow us to quantify which waves dissipate due to radiative damping, and which 
successfully convert into shocks.  Returning to equation  (\ref{eq:DiffusiveLoss_General2}), we can define at each radius the critical frequency for the wave that damps by radiative diffusion in a single radiation pressure scale height: 
\begin{equation}
\omegarad^2 \simeq 2\gamma_0(\gamma_0-1) \frac{c_{s0}^2}{\Hprad^2}  \frac{\Lwmax}{L_{\rm rad}}; 
    \label{eq:DiffusiveLoss_General2}
\end{equation}
the instantaneous damping time due to diffusion is $t_{\rm damp}\equiv {L_w}/{|\dot L_w|} =  (\Hprad/ c_{s0})\omegarad^2/\omega^2$.  We can also write the shock formation criterion $\int_{t_i}^t dt'/t_{\rm shock} = 1$ where $t_{\rm shock} = [2/(\gamma_0+1)] (\Lwmax/L_w) \omega^{-1}$.   Evaluating the ratio of diffusion time to shock time at frequency $\omega=\omegarad$, 
\begin{equation}
    \left. t_{\rm damp}^2\over t_{\rm shock}^2\right|_{\omega =\omegarad } = 
    {\gamma_0 (\gamma_0-1)(\gamma_0+1)^2 \over 2} {L_w\over L_{\rm rad}}. 
\end{equation}
This analysis indicates a shock forms before radiative damping has had time to act, for any waves more luminous than the stellar radiative luminosity.  Therefore, {\em  wave-driven outbursts that exceed the envelope Eddington luminosity must involve shock formation. }  This is especially true for super-Eddington outbursts, as the quiescent luminosity is usually well below the envelope's Eddington limit. 

\begin{figure*}
	\centering
    \includegraphics[width=0.95\textwidth]{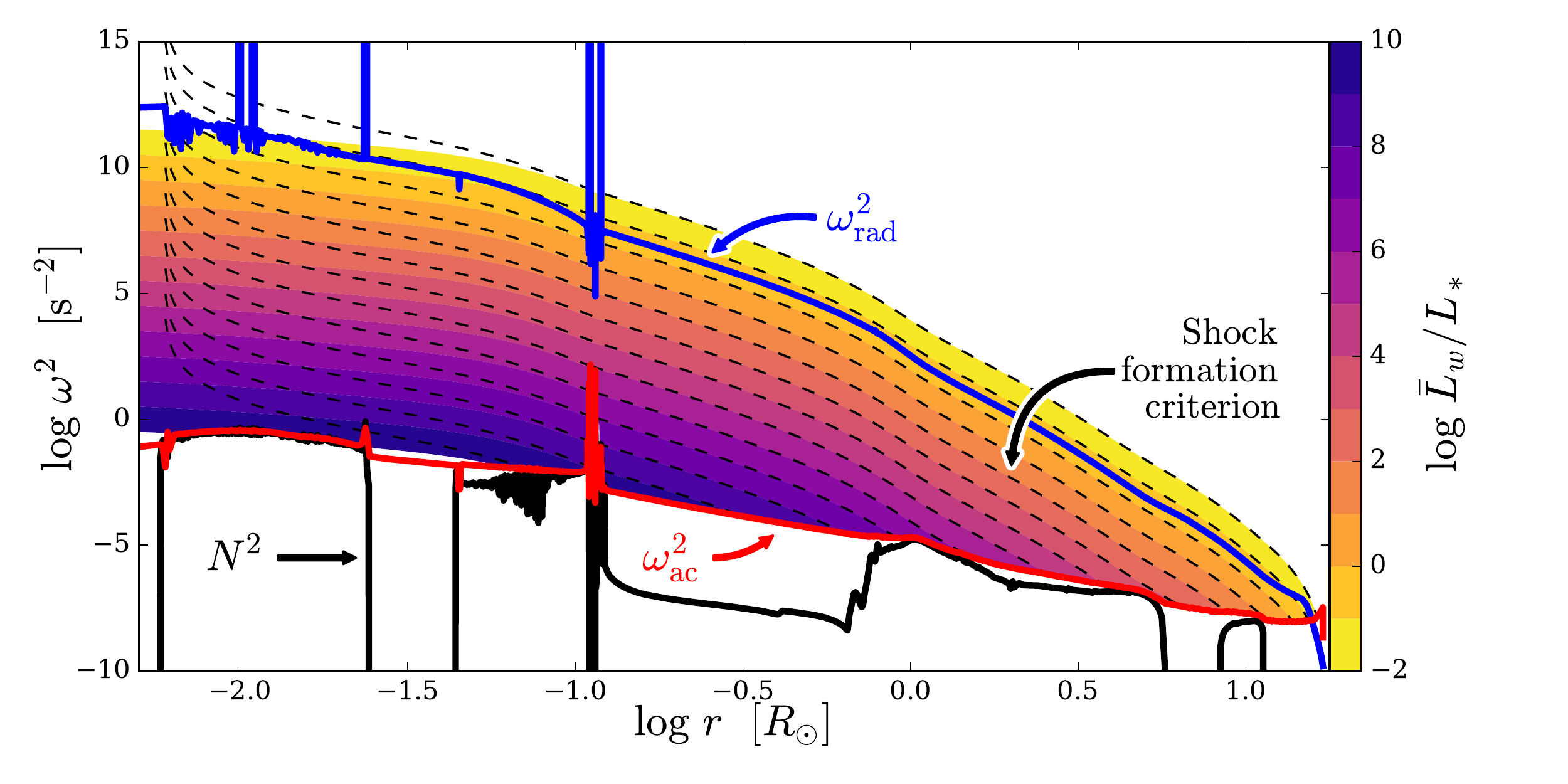}
    \caption{Wave propagation diagram of a core oxygen-burning blue supergiant model evolved in MESA. Contours show the exact (eq. \ref{eq:whereWaveShocks}) shock formation locations for acoustic waves with frequency $\omega$ and peak luminosity $\bar L_w$ launched from the stellar center $r_i=0$ (filled contours) or from the convective boundary $r_i=10^{-2.2}R_\odot$ (dashed lines).  Note that these are indistinguishable in the outer envelope.  Also plotted are the Brunt-V\"ais\"al\"a ($N^2$), acoustic cutoff ($\omega_{\rm ac}^2$), and radiative damping ($\omega_{\rm rad}^2$) frequencies. Waves with luminosities $\bar L_w>[8 \gamma_0(\gamma_0-1) ]^{-1}L_{\rm rad}\simeq L_{\rm rad}/5$ produce shocks rather than damping by radiative diffusion. 
    }
    \label{fig:propagation_diagram}
\end{figure*}

In figure (\ref{fig:propagation_diagram}) we examine acoustic propagation, radiative damping, and shock formation within a model star  generated by the MESA code (r8118)  from an initial solar-metallicity object of 50 $M_{\odot}$.  At the time of the figure, stellar winds have removed all but $22\ M_{\odot}$ and core oxygen burning has just begun.   We plot the Brunt-V\"ais\"al\"a ($N$), acoustic cutoff ($\omega_{\rm ac}$), and radiative damping frequencies ($\omegarad$), as well as the shock formation radius (eq.\ \ref{eq:whereWaveShocks}). As is clear from the figure, shock formation outpaces radiative damping for $L_w>10^{5.2} L_\odot $, somewhat below the stellar luminosity of $10^{5.9} L_\odot$.

\section{Discussion} \label{S:Discussion}

In this work we have found that strong stellar outbursts and wave-driven outflows necessarily involve shock formation, rather than radiative dissipation.  The distinction is important because shock formation occurs at a different radius and deposits wave energy in a distributed fashion, and can lead to mass ejection at the surface.  

Moreover, we find that the condition for shock formation within stars can be predicted with a single expression (equation \ref{eq:whereWaveShocks}).  This result is remarkably general, as it applies to any one-dimensional motion within a non-isentropic fluid; yet it is almost as simple as the classic result for planar, isentropic flows.  Our exact derivation, which we obtained by generalizing an analysis from the sonoluminescence literature \citep{FLM:68109}, matches a heuristic calculation of wave crossings that using wave luminosity conservation (which can be considered a consequence of the wave action principle due to \citealt{1970PhFl...13.2710D}, in the absence of reflections).      

Because of its generality, our result applies equally well to a single wave pulse and to a continuous wave.   Furthermore, the formation of a shock from one wave  tends to be unaffected by the passage of earlier waves, because these typically deposit energy and momentum only after they have steepened into shocks.    Therefore our criterion for shock formation remains reasonably valid even within stars that have been set into motion by a strong pulse or a steady acoustical flux. 

Our next task is to predict in detail the propagation of weak shocks outside the shock formation radius, in order to understand the transition from weak to strong and to determine the patterns of heat and momentum deposition.  We shall address this in subsequent papers. 

\acknowledgments SR acknowledges support from a Gilchrist Fellowship.  SR and CDM are supported by a Discovery Grant from NSERC, the Canadian National Sciences and Engineering Research Council.  CDM is very grateful to the members of the Monash Centre for Astrophysics for support, hospitality, and stimulating discussions.

\appendix
\section{Taylor expansion around a wave node}
Substituting the Taylor expanded fluid variables into the fluid equations generates the following collection of $\xi^0$ and $\xi^1$ order terms. Note that we simplify the notation by substituting $a\equiv c_s$. 

\begin{align}
        a_0' &= a_1 +\frac{\gamma_0}{2}u_1 \Label{a1} & & \quad \quad\quad \quad \quad \quad\quad \quad\quad \quad\quad \quad\mathrm{N/A} \\
        a_0u_1 &= \frac{a_0^2}{\gamma_0}\frac{p_1}{p_0} + g \Label{a2} & 0&= (-a_0u_1+g)\left( \frac{\gamma_1}{\gamma_0}+\frac{p_1}{p_0} \right) + u_1'-a_0u_2+u_1^2 + \frac{2a_0a_1}{\gamma_0}\frac{p_1}{p_0}+ \frac{a_0^2}{\gamma_0}\frac{p_2}{p_0}, \\
        a_0\frac{p_0'}{p_0} &= a_0\frac{p_1}{p_0} - \gamma_0u_1  \Label{a3} & 0 &= \frac{p_1'}{p_0} - a_0\frac{p_2}{p_0} +\gamma_1u_1 + \gamma_0\left(u_2 + \frac{\alpha u_1}{r} \right) + (\gamma_0+1)u_1\frac{p_1}{p_0} \\       
    \gamma_0u_1 &= a_0(\gamma_0' - \gamma_1)   \Label{a4} & \gamma_1' &=  \gamma_0u_2 + \gamma_2a_0 
\end{align}
We leave out the $\xi^1$ result from the continuity equation (A2) since it is egregious in length and not used in the next derivation. The variables that emerge in (A2) include $\gamma_0$, $u_1$, $u_2$, $a_0$, $a_1$, $a_2$, and $a_1'$.

\section{Derivation }
The goal is to find a final differential equation with $u_1$, $u_1'$ and any quiescent variables (subscript 0). Through a process of elimination using equations (A3), (A4), (A6), and the derivative of (A3), one can generate the following differential equation:
\begin{equation}
    0= 2u_1' + (\gamma_0+1) u_1^2 + \left(a_0' + \frac{\gamma_0'}{\gamma_0}a_0 + \frac{\alpha a_0}{r} - \frac{\gamma_0 g}{a_0}\right)u_1 .
\end{equation}
The quiescent gas is initially in hydrostatic equilibrium, which satisfies $p_0'=-g\rho_0$. Since the local quiescent sound speed is $a_0^2=\gamma_0p_0/\rho_0$, we can eliminate the body force (i.e., gravity) by substituting $-\gamma_0 g /a_0 = a_0 p_0'/p_0$.

With respect to $u_1(t)=y(t)$, this equation is known as the Bernoulli equation $y' + p(t)y + q(t)y^n=0$ for $n=2$. The solution can be written explicitly in the form 
\beq
u_1^{-1}(t) =e^{-\phi(t)}\left( u_1^{-1}(0) +  \int_0^t \left(\frac{\gamma_0+1 }{2}\right)e^{\phi(\tau)}d\tau  \right), 
\eeq
where,
\beq
\phi(t) \equiv-\frac{1}{2}\int_0^t\left(a_0' + \frac{\gamma_0'}{\gamma_0}a_0 + \frac{\alpha a_0}{r} - \frac{a_0p_0'}{p_0}\right)d\tau.  \nonumber
\eeq
We can rewrite the integrand in terms of the radius, since $d\tau=dr/a_0$, and integrate the expression to obtain
\bea
-2\phi(r) &\equiv& \int\left(\frac{a_0'}{a_0} + \frac{\gamma_0'}{\gamma_0} + \frac{\alpha }{r} - \frac{p_0'}{p_0}\right)dr = \mrm{ln} \left( \gamma_0 r^{\alpha }a_0p_0\right)\bigg|_{r_0}^r = -\mrm{ln} \left( r^{\alpha }\rho_0a_0^3\right)\bigg|_{r_0}^r \implies e^{\phi(r)} = \sqrt{\frac{L_{\mrm{max}} (r_0)  }{L_{\mrm{max}} (r) } }.
\eea
Thus, the wave front evolution can be solved analytically with the following expression
\beq
u_1^{-1}(r) = \sqrt{\frac{L_{\mrm{max}} (r)  }{L_{\mrm{max}} (r_0) } }\left( u_1^{-1}(r_0) + \int_{r_0}^r  \left(\frac{\gamma_0+1 }{2}\right)\sqrt{\frac{L_{\mrm{max}} (r_0)  }{L_{\mrm{max}} (\tilde{r}) } }\frac{d\tilde{r}}{a_0}  \right). 
\nonumber
\eeq
Notice that the body force $g$ is never defined explicitly. Since $g$ is arbitrary, the analytic result must be valid for an arbitrary distribution of fluid. It is also applicable for waves in planar, cylindrical, and spherical symmetry ($\alpha=$ 0, 1, 2).

\end{document}